\documentclass[preprint,prb,showpacs,preprintnumbers,amsmath,amssymb]{revtex4}

\usepackage{graphicx}
\usepackage{dcolumn}
\usepackage{bm}

\usepackage{graphicx}
\usepackage{dcolumn}
\usepackage{bm}

\begin{document}




\title{Magnetic and electrical characterization of lanthanum superconductors with dilute Lu and Pr impurities.}

\author{F\'elix Soto} 
\author{Luc\'ia Cabo}
\author{Jes\'us Mosqueira}
\author{F\'elix Vidal}
 \email{fmvidal@usc.es}

\affiliation{Laboratorio de Baixas Temperaturas e Superconductividade (Unidad Asociada al ICMM-CSIC, Spain), Departamento de F\'\i sica da Materia Condensada, Universidade de Santiago de Compostela, E-15782 Spain \mbox{}\vspace{2cm}\mbox{}}


\begin{abstract}
In this work we present measurements of the electrical conductivity and of the magnetization of La and of La-Pr and La-Lu dilute (up to \mbox{2 at.$\,$\%}) alloys, from which we determine, very in particular, the influence of dilute magnetic impurities on the upper critical magnetic field amplitude [and then on the superconducting coherence length amplitude $\xi(0)$] and on the Ginzburg-Landau parameter~$\kappa$.
 \end{abstract}


\maketitle


As it is well known since the pioneering experiments of Matthias and coworkers\cite{Matthias1,Matthias2}, the La-Pr and La-Lu alloys are very well suited to study the different aspects of the interplay of superconductivity with magnetic impurities.\cite{Fisher} However, even today some of the main superconducting characteristics of these alloys are still not well settled or even they have not been measured at all. This is, for instance, the case of a so-central parameter as the Ginzburg-Landau superconducting coherence length amplitude, $\xi(0)$. Here we will present detailed magnetic and electrical measurements in La and La-Pr and La-Lu dilute (up to \mbox{2 at.$\,$\%}) alloys, from which we will obtain, very in particular, $\xi(0)$ and the Ginzburg-Landau parameter, $\kappa$.

All the samples (pure La and La-Pr and La-Lu alloys) used in this work are commercial (Goodfellow and Alfa Aesar, 99.9\% purity). The manufacturers warrant that the impurity concentrations are below 100~ppm (by weight) for any of the following ions: Al, Ca, Fe, Mg, Nd, Ni, Pr, Si, and Y. X-ray diffraction analyses revealed that the pure La samples consist in a randomly-oriented mixture of two crystallographic phases, double-hexagonal-close-packed ($\alpha$-La) and face-centered-cubic ($\beta$-La), which is a common feature in most of the La samples studied in other works.\cite{Leslie,Anderson,Berman,Finnemore,Johnson,Legvold,Pan}  For the alloys, we indicate in the last column in Table~I   the average distance between Pr or Lu impurities, $d_{imp}$, obtained from the samples' densities and nominal compositions. The values are between 12 and 19~${\rm \AA}$.

A first magnetic characterization of the different samples has been done through measurements of the field-cooled magnetic susceptibility versus temperature curves, $\chi^{FC}(T)$. Some examples of the obtained curves are shown in Fig.~1(a). These measurements were performed with a high resolution, superconducting quantum interference device (SQUID) based, magnetometer (Quantum Design's MPMS). For that, each sample was first cut as a cylinder of 6~mm height and 6.5~mm diameter, and care was taken just before each measurement of removing the possible superficial insulating oxide layer with sandpaper (Buehler, 600 grit) and then immersing the samples in an acetone ultrasonic bath for several minutes. The external magnetic field applied in the measurements was $\mu_0H = 0.5$~mT, much smaller than the lower critical magnetic field amplitude (see Refs.~[\onlinecite{Leslie,Anderson,Berman,Finnemore,Johnson,Legvold,Pan}] and also below). For each sample, the resulting magnetic susceptibility was corrected for demagnetizing effects through the ellipsoidal approximation and then normalized to the ideal value -1 at low temperatures. Other details of these last measurements are similar to those we have followed in magnetization experiments in other superconductors.\cite{Mosqueira} In the Fig.~2(b) the solid lines are fits to the experimental $d\chi^{FC}/dT$ points of a Gaussian distribution, $(\sqrt{(\pi/2)} \Delta T_{C0})^{-1}{\rm exp}[-2(T-T_{C0})^2/\Delta T_{C0}^{2}]$ with $T_{C0}$ and $\Delta T_{C0}$ as free parameters.   We will see below that these $T_{C0}$ temperatures, that are summarized in Table~I, are consistent with the transition temperatures that may be obtained from resistivity versus temperature curves. The $\Delta T_{C0}$ values, which are also summarized in Table~I, provide a good indication of how much the structural and stoichiometric inhomogeneities at long length scales [well larger than $\xi(0)$] affect the normal-superconducting transition temperature of each sample. It is also worth noting here that in the case of the nominally pure La samples their $T_{C0}$ values are well between the ones for pure $\alpha$-La ($\sim 5.0$ K) and $\beta$-La ($\sim 6.0$ K), \cite{Leslie,Anderson,Berman,Finnemore,Johnson,Legvold,Pan} but no traces were observed of diamagnetic transitions near these last temperatures (which would manifest themselves as steps in the $\chi^{FC}$ curve). This is consistent with a mixing of both crystallographic phases at lengths of the order or smaller than $\xi(0)$.\cite{Anderson}  We note also that the $T_{C0}$ values of Table~I are consistent with both the experimental results of Matthias and coworkers\cite{Matthias1,Matthias2} and the Abrikosov-Gor'kov predictions\cite{abrikosov}. These last aspects will be analyzed in detail elsewhere.

In Fig.~\ref{susceptibilidad} we present measurements, corresponding to three of the superconducting compound  studied in this work, of the magnetization under a constant magnetic field of $\mu_0H=50$~mT and including temperatures  well above the superconducting transition. These measurements may be useful, {\it e.g.}, to analyze the superconducting fluctuations above $T_{C}$ in these samples, as it will be reported elsewhere. Note that the absolute values of $M(T)$ above the transition increase with the Pr content but, for instance, in the La-1~at.\%Pr alloy the magnetization at 1.5$T_C$ is only around 5 times larger than the one of the optimally-doped YBa$_2$Cu$_3$O$_{7-\delta}$.\cite{medidasYBCO} As for the $T$-dependence of these magnetization curves, they present a Curie-like curvature, ressembling the one of 
optimally-doped YBa$_2$Cu$_3$O$_{7-\delta}$. 
We have also performed measurements of the magnetization versus applied magnetic field at a constant temperature above the transition; however, the understanding of the resulting $M(H)_T$ curves is complicated by the existence of an upturn at low fields, perhaps related to the presence of small amounts of inhomogeneities which do not affect the $M(T)_H$ dependence.\cite{Lucia}

To perform measurements of the resistivity against temperature, thin slabs (typically $5\times1\times0.1\:{\rm mm^3}$) were cut from the same rods used in the magnetization experiments. The electrical contacts were made by attaching to the samples four Al-Si wires ($25 \mu {\rm m}$ diameter) with an in-line geometry by using an ultrasonic micro-wire bonder (Kulicke $\&$ Soffa, model 4523). Just before attaching the contacts the possible superficial insulating oxide layer was again removed with sandpaper  and   an acetone ultrasonic bath. The final resistance was typically 50~m$\Omega$ per contact. The ac (34 Hz) current supplied to measure the samples' resistance was 2 mA, and the longitudinal voltage was measured by using a lock-in amplifier (EG$\&$G Princeton Applied Research, model 5210) which has a resolution of 0.05\% up to 1 nV. The main uncertainty in these measurements comes from the samples' dimensions and is around 15\%. A general view of the resulting resistivity versus temperature curves is presented in Fig.~2 for some of the samples. The nominally pure La samples have a room-temperature resistivity ($\sim 50\:\mu\Omega$cm) and a temperature dependence comparable to the one of pure $\alpha$-La and $\beta$-La samples.\cite{Legvold} Note also that the room-temperature resistivity increases monotonically with the Pr concentration. The inset of Fig.~2 presents a detail of these measurements around the superconducting transition. By taking into account that the resistivity falls to zero at the temperature at which a superconducting path exists through the sample ({\it i.e.}, when the superconducting volume fraction exceeds $\sim 15 \%$ \cite{Kirkpatrick}) the transition temperature obtained from those measurements are in good agreement with the ones obtained from the low-field $\chi^{FC}$ measurements presented in Fig.~1. The values of $\ell$, the normal-state electronic mean field path extrapolated to  $T=0$~K, were estimated from the residual resistivities $\rho^{res}$ through a simple Drude-model relation,\cite{Ashcroft} and compiled in Table~I. By comparing these $\ell$ values with the corresponding $\xi(0)$ (see below), one may conclude that all the samples are in the dirty limit. In the case of nominally pure La, this is consistent with the presence in these samples of a small amount of impurities. The relaxation time of normal electrons corresponding to those $\ell$ values are of the order of $\tau\sim 10^{-14}$ s.

Other basic superconducting parameters were obtained from measurements of the magnetization against magnetic field at different constant temperatures below the superconducting transition. Some examples of these measurements are shown in Fig.~4. These curves are, even in the case of the pure La, typical of type II superconductors, in agreement with the earlier results of Pan and coworkers\cite{Pan} which showed that even high-purity $\alpha$-La is a type II superconductor with a Ginzburg-Landau parameter $\kappa\approx2.4$. For all the compounds studied here we found that the mixed-state magnetization is highly irreversible up to the upper critical field, $H_{c2}(T)$. To overcome this difficulty in determining $H_{c2}(T)$, the reversible magnetization was approximated as $M_{rev}=(M^++M^-)/2$, where $M^+$ and $M^-$ are the magnetization when increasing and decreasing the external magnetic field respectively. The resulting $M_{rev}(H)_T$ curves present a linear behaviour over a wide region near $H_{c2}(T)$, consistently with the Abrikosov prediction for the magnetization in the high-field regime,\cite{Tinkham}
\begin{equation}
\label{Mrev}
M_{\rm rev} = \frac{H-H_{c2}(T)}{\beta_A \left[2\kappa^2(T)-1\right]},
\end{equation}
with $\beta_A\approx 1.16$ for a triangular vortex lattice. For each sample, the solid straight lines in Fig.~4 are the best fits of Eq.~(\ref{Mrev}) to the $M_{rev}(H)_T$ curves in the linear region, with $H_{c2}(T)$ and $\kappa (T)$ as free parameters (for each temperature). The resulting $H_{c2}(T)$, presented in the Fig.~5(b), show a linear temperature dependence in the interval studied ($0.75<T/T_{C0}<0.95$). This allows the obtainment of $\xi(0)$ through the relationship\cite{Tinkham} $-\mu_0(dH_{c2}/dT)_{T_{C0}}=\phi_0/2\pi T_{C0}\xi^2(0)$. The uncertainties in the so-obtained $\xi(0)$ values are of the order of 20\%.  We summarize in Table~I these results for the $\xi(0)$ of each sample. The results for $\kappa(T)$ are presented in Fig.~5(a). They are almost temperature independent, and linear extrapolations to $T_{C0}$ give the values of $\kappa(T_{C0})$ summarized also in Table~I. Note that in the case of pure La, these values are compatible with the ones that may be found in the literature. For instance, in Refs.~[\onlinecite{Anderson,Finnemore,Johnson,Pan}] is proposed a thermodynamic critical magnetic field $\mu_0H_c(0)\sim80-84$ mT for pure $\alpha$-La and $110-160$ mT for $\beta$-La. By using the relation $\mu_0H_c(0)=\mu_0H_{c2}(0)/\sqrt{2}\kappa$ and the $\kappa$ values obtained from our magnetization curves, we get $\mu_0H_c(0)\approx95$ mT. We are not aware of the existence of previous measurements of $\kappa$ or $H_{c2}(0)$ in the La-Pr and La-Lu alloys studied here. In Fig.~6 it is shown the dependence with the impurity concentration of the obtained $\kappa(T_{C0})$ and $H_{c2}(0)$ values for all the samples studied. As may be seen, $\kappa$ scarcely varies with the concentration of nonmagnetic impurities but it increases slightly with the amount of Pr. $H_{c2}(0)$ seems to increase a little with the Pr concentration.

In conclusion, in this work we have obtained the $\kappa(T_C)$, $\xi(0)$ and $\ell$ values for pure La and, for the first time, La-Pr and La-Lu dilute (up to \mbox{2 at.$\,$\%})  alloys.  The resulting values are summarized in Table~I. These superconducting parameters have a considerable interest, {\it e.g.}, for further studies we are performing about the behaviour of the fluctuating Cooper pairs in presence of magnetic impurities.

\section*{Acknowledgements}
This work was supported by the CICYT, Spain (grants no.\ MAT2004-04364), the Xunta de Galicia (PGIDT01PXI20609PR), and Uni\'on Fenosa (220/0085-2002). LC acknowledges financial support from Spain's Ministerio de Educaci\'on y Ciencia through a FPU grant.

\newpage

\begin{table}
\begin{center}
\caption{\label{tab:Table} Summary of the main superconducting parameters of the La-Pr alloys studied in this work. For comparison, we show also data for one of the pure La samples and for one of the La-Lu alloys. They were obtained from the magnetization and electrical resistivity measurements described in the main text. In the last column it is shown an estimation of the average distance between impurities, $d_{imp}$, obtained from the samples' nominal compositions.\\ \mbox{}\\}
\vspace{10 pt}
\begin{tabular}{lccccccc}
\hline
\hline
Sample  \mbox{}\hspace{3em}\mbox{}  &
  \mbox{}\hspace{0.5em}\mbox{} $T_{C0}$ \mbox{}\hspace{0.5em}\mbox{}&
 \mbox{}\hspace{0.5em}\mbox{}  $\Delta T_{C0}$ \mbox{}\hspace{0.5em}\mbox{}&
 \mbox{}\hspace{0.5em}\mbox{}   $\mu_0H_{C2}(0)$ \mbox{}\hspace{0.5em}\mbox{}&
  \mbox{}\hspace{0.5em}\mbox{}   $\xi(0)$ \mbox{}\hspace{0.5em}\mbox{}&
   \mbox{}\hspace{0.5em}\mbox{}   $\kappa(T_{C0})$ \mbox{}\hspace{0.5em}\mbox{}&
   \mbox{}\hspace{0.5em}\mbox{}   $\ell$ \mbox{}\hspace{0.5em}\mbox{}&
   \mbox{}\hspace{0.5em}\mbox{}    $d_{imp}$ \mbox{}\hspace{0.5em}\mbox{}\\ 
 & (K)& (K)& (T)& 
 (\AA)&&(\AA)&(\AA)\\ \hline

La    &5.85&0.16&0.8&200&4.4&265& \\ 
La-0.5 at.\% Pr  &5.69&0.12&0.9&190&5.6&150&19 \\
La-1 at.\% Pr&5.51&0.25&0.9&180&6.0&110&15\\ 
La-2 at.\% Pr&5.40&0.18&1.0&180&6.4&75&12\\
La-2 at.\% Lu&5.88&0.26&0.9&190&4.2&210&12\\
\hline
\end{tabular}
\end{center}
\end{table}

\newpage

\begin{figure}
\begin{center}
\includegraphics[width=0.7\textwidth]{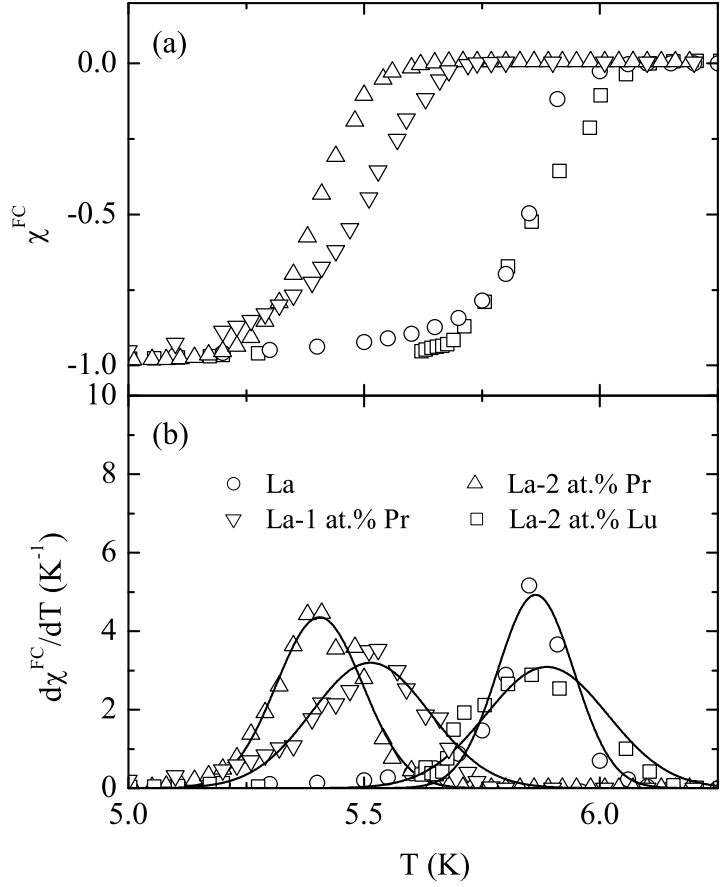}
\end{center}
\caption{(a) Some examples of our measurements at low fields ($\mu_0H = 0.5$ mT) of the field-cooled magnetic susceptibility versus temperature. These data were corrected for demagnetizing effects and normalized to the ideal value -1 at low temperatures. (b) The corresponding $T_C$ distributions obtained as the temperature derivative of the normalized $\chi^{FC}$ curves. The solid lines are fits of a Gaussian distribution with $T_{C0}$ and $\Delta T_{C0}$ as free parameters (see main text for details).}
\label{teces}
\end{figure}

\newpage

\begin{figure}[b]
\begin{center}
\includegraphics[width=0.55\textwidth]{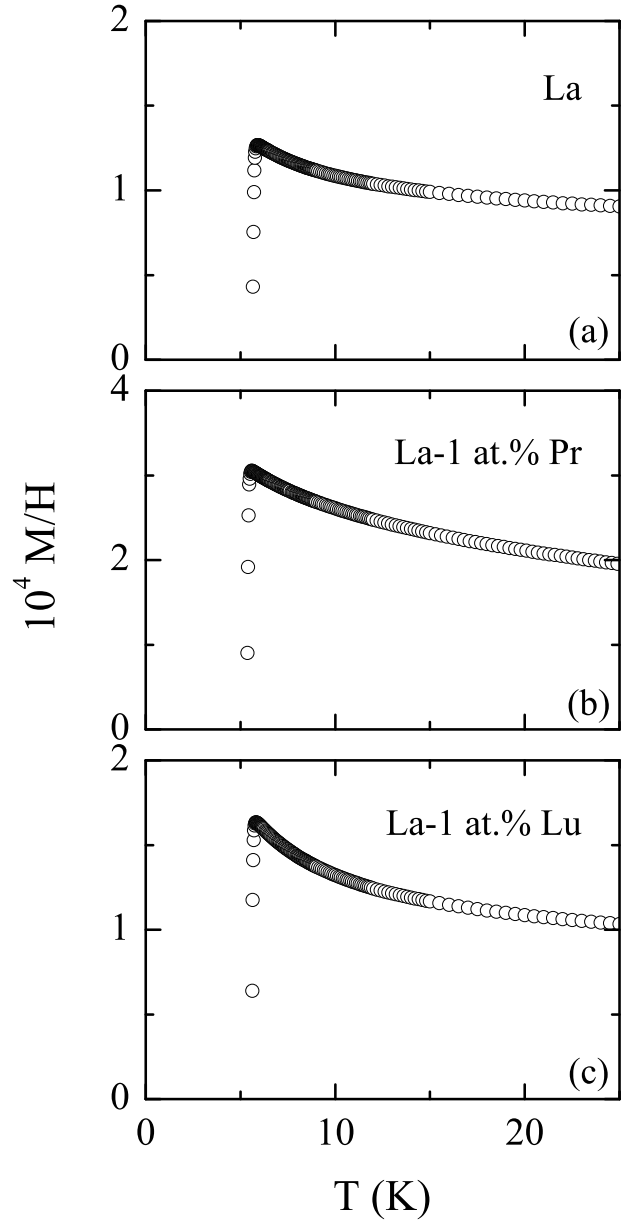}
\end{center}
\caption{Some examples of the temperature dependence of $M/H$ measured under low magnetic fields (in these examples with $\mu_0H=50$~mT) for three of the superconducting compounds studied in this work.}
\label{susceptibilidad}
\end{figure}

\newpage

\begin{figure}[b]
\begin{center}
\includegraphics[width=0.9\textwidth]{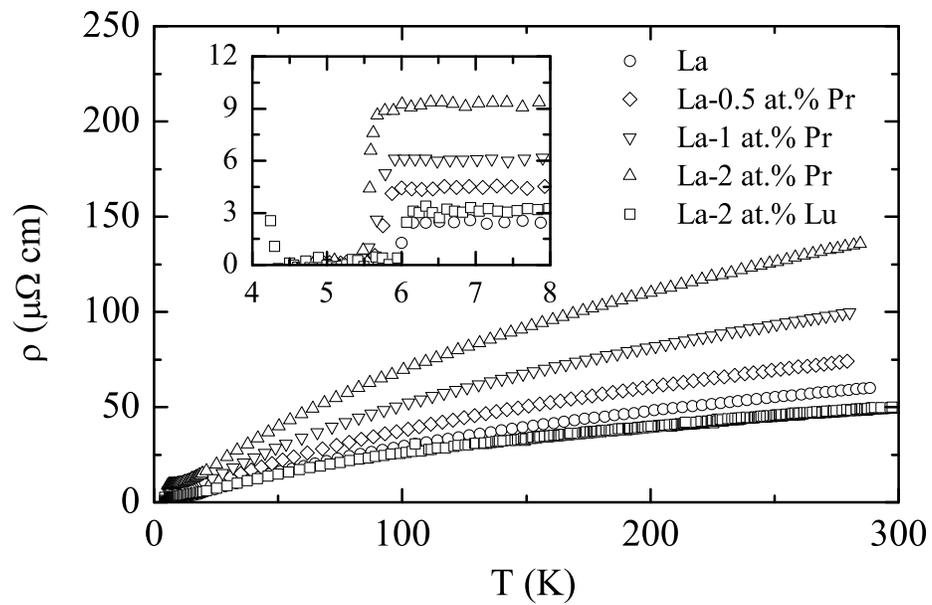}
\end{center}
\caption{Temperature dependence of the electrical resistivity of some of the samples studied. Inset: Detail around the superconducting transition.}
\label{resistividad}
\end{figure}

\newpage

\begin{figure}[b]
\begin{center}
\includegraphics[width=1.05\textwidth]{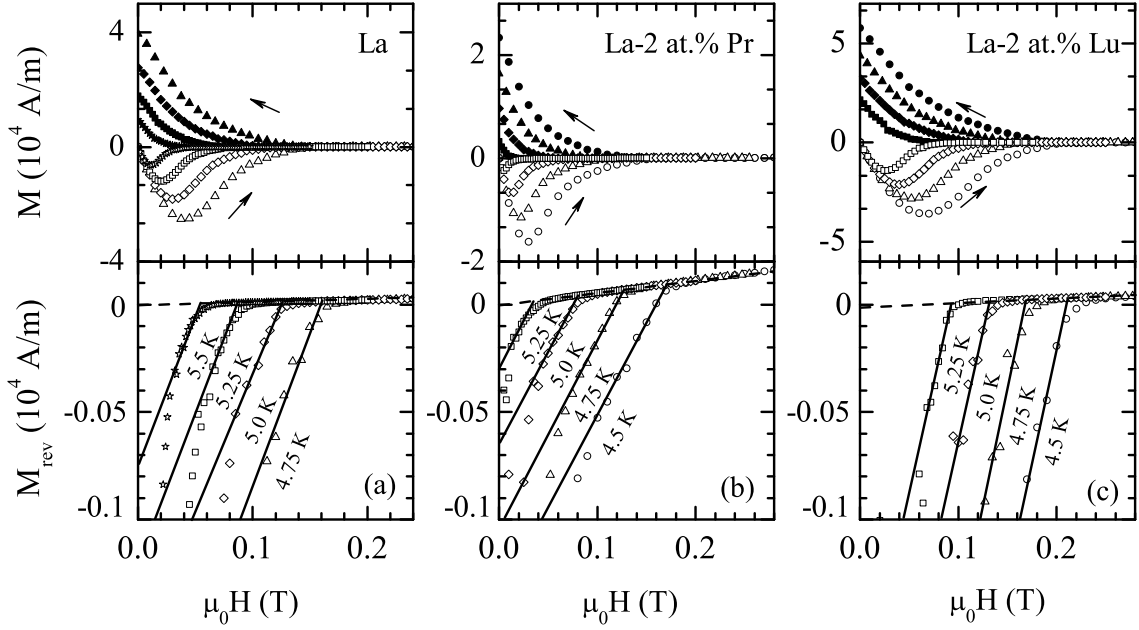}
\end{center}
\caption{Upper panels: Examples corresponding to some of the samples studied of the magnetization vs. applied field at some constant temperatures below $T_{C0}$. They were obtained by increasing (open symbols) and then decreasing (closed symbols) the applied magnetic field. Lower panels: The reversible magnetization curves obtained as the semisum of field-up $M^+$ and field-down $M^-$ magnetizations. The dashed lines correspond to linear fits to the data above $H_{c2}(T)$. The solid lines are fits of the Abrikosov theory for the mixed-state magnetization to the data in the linear region, with $H_{c2}(T)$ and $\kappa(T)$ as free parameters (see main text for details).}
\label{MvsH}
\end{figure}

\newpage

\begin{figure}[b]
\begin{center}
\includegraphics[width=0.8\textwidth]{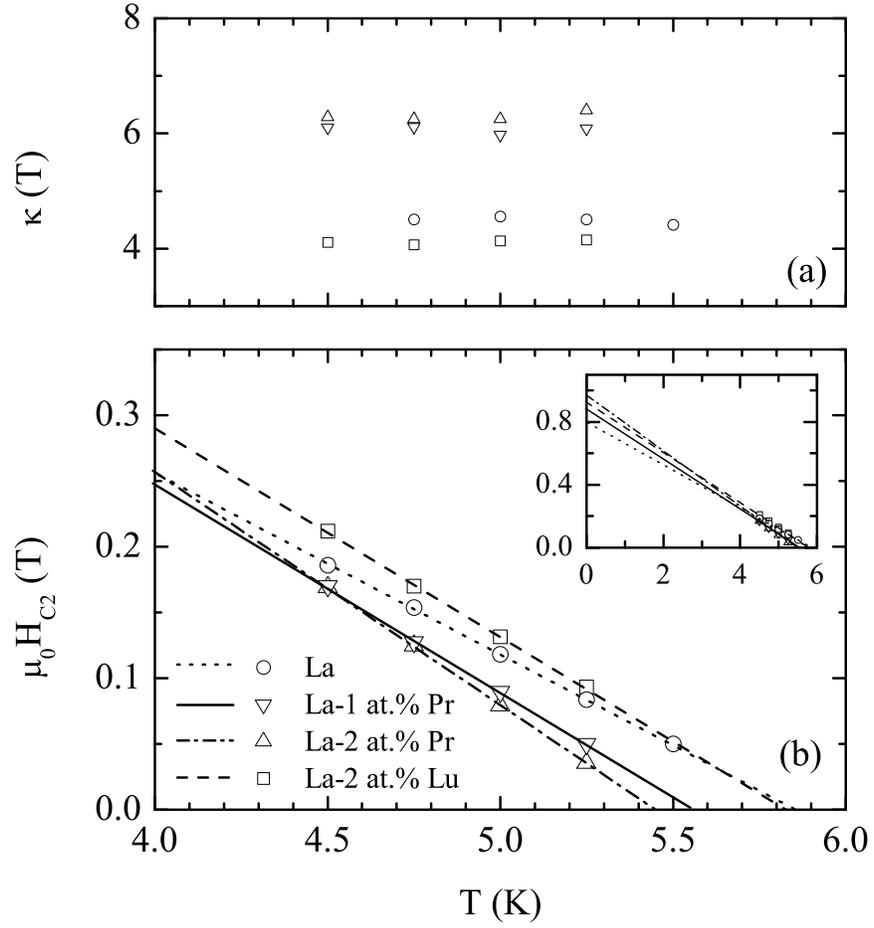}
\end{center}
\caption{Some examples of the temperature dependences of (a) the Ginzburg-Landau parameter and (b) the upper critical magnetic field. These quantities were obtained from $M(H)_T$ measurements like the ones shown in Fig.~4. In the inset of Fig.~5(b) it is presented a linear extrapolation of $\mu_0H_{c2}(T)$ to $T=0$~K.}
\label{HvsT}
\end{figure}

\newpage

\begin{figure}[b]
\begin{center}
\includegraphics[width=0.7\textwidth]{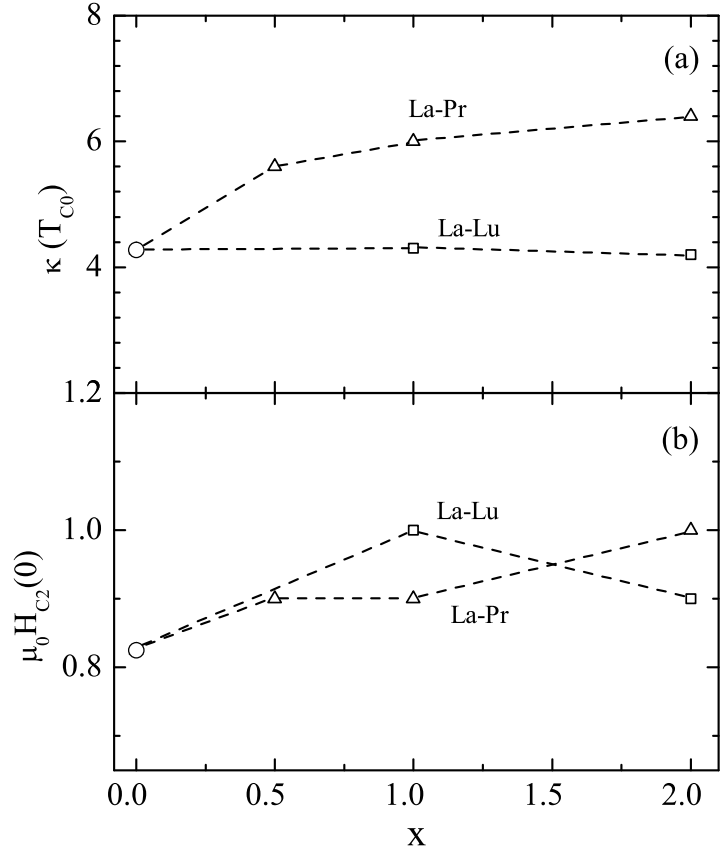}
\end{center}
\caption{Dependence with the nominal amount of impurities of (a) the Ginzburg-Landau parameter at $T_{C0}$ and (b) the Ginzburg-Landau amplitude of the upper critical magnetic field.}
\label{dependx}
\end{figure}

\end{document}